\documentclass[aps,print,showpacs,onecolumn]{revtex4}%
\usepackage{amsfonts}
\usepackage{amsmath}
\usepackage{amssymb}
\usepackage{graphicx}%
\begin{document}
\author{Gang Chen$^{a,b\ },$\ J.-Q. Liang$^{a}$, W.-M. Liu$^{c}$}
\affiliation{$^{a}$Institute of Theoretical Physics, Shanxi University, Taiyuan 030006, China}
\affiliation{$^{b}$Department of physics, Shaoxing College of Arts and Sciences, Shaoxing
312000, China}
\affiliation{$^{c}$Beijing National Laboratory for Condensed Matter Physics, Insittute of
Physics, Chinese Academy of Sciences, Beijing 100080, China}
\title{Quantum Phase Transition in{\LARGE \ }Ultracold $^{87}$Rb Atom Gas with
Radiant Field}

\begin{abstract}
A second-order quantum phase transition in two-species Bose-Einstein
condensates of $^{87}$Rb atoms coupled by a quantized radiant field is
revealed explicitly in terms of the energy spectrum which is obtained in the
thermodynamic limit $1/N\rightarrow0$ and is controllable by the coupling
parameter between the atom and field. The scaling behavior of the collective
excitation modes at the critical transition point is seen to be in the same
universality class as that of the Dicke model. It is also demonstrated that
the quantum phase transition is realizable below the critical temperature of
BEC and can be detected experimentally by measuring the abrupt change of atom
population imbalance.

\end{abstract}

\maketitle

Experimental realization of Bose-Einstein condensate (BEC) in trapped atomic
gases has opened a possibility to explore the quantum mechanics of mesoscopic
systems in a quantitatively new regime \cite{1,2,3}. A paradigmatic effect in
such systems is the coherent exchange of particles between weakly connected
states, which was first predicted in superconductors \cite{4} known as
Josephson effect and is expected to be displayed also in trapped BECs by a
double-well potential \cite{5,6,7} or optical lattice \cite{8,9,10}. In
particular, recent experiments on two-component BECs of $^{87}$Rb atoms have
stimulated considerable interests in the phase dynamics and number
fluctuations of the condensates \cite{11,12} in a radiant field which induces
the coherent exchange of particles between two-species BECs. It has been
demonstrated that the population oscillation between two components modulated
by the collapses and revivals indicates the quantum nature of the system
\cite{13,14}. The quantum system of two-level atoms coupled by a quantized
field exhibits a novel phenomenon of super-radiance describing the collective
emission of light from excited atoms and involving the spontaneous buildup of
coherence in a macroscopic ensemble of atoms. Recently, great attentions have
been attracted on super-radiant scattering of BECs consisting of atoms with
two internal states both experimentally \cite{15,16}and theoretically
\cite{17,18}.

At zero-temperature all thermal fluctuations are frozen out whereas quantum
fluctuations are dominative. The microscopic quantum fluctuations can give
rise to a macroscopic phase transition in the ground state of many-body
systems associated with the variation of a controllable parameter \cite{19}
which is now known as the quantum phase transition (QPT) originating from
singularity of the energy spectrum \cite{20}. Trapped BECs, which are well
isolated from the environment, can be excited either by deforming the trap or
by radiant field and thus are ideal quantum systems to observe the QPT. It has
been shown that ultracold bosonic atoms loaded in an optical lattice possess a
continuous quantum phase transition from a superfluid phase to a Mott
insulator phase at low temperatures induced by varying the depth of the
optical potential \cite{21,22}. While the super-radiant phase transition of
BECs consisting of atoms with two internal states has not yet been
investigated and may be of more fundamental importance.

In this letter we reveal a novel second-order quantum phase transition in the
two-component BECs coupled by a radiant field with Raman coupling, which
belongs to the same universality class as in the Dicke model describing the
collective effects of $N$ two-level atoms interacting with a single photon
mode in quantum optics \cite{23,24,25} where the atomic ensemble in the normal
phase is collectively unexcited while is macroscopically excited with coherent
radiations in a so-called super-radiant phase. As a matter of fact what we
considered here is a generalized Dicke Hamiltonian with atom-atom interaction
in addition. \textbf{\ }We obtain the ground state energy as a function of the
coupling parameter between the atom and field which possesses a critical point
at which the system undergoes a second-order phase transition. The QPT and the
scaling behavior at the transition point are demonstrated explicitly by
variation of the coupling parameter. Finally we evaluate the critical value of
coupling parameter at phase transition point to show that the energy scale of
it is realizable below the critical temperature of BEC and the super-radiant
phase transition can be detected experimentally by measuring the atom
population imbalance between two species.

We consider the ultracold $^{87}$Rb atom gas consisting of two internal states
$\left|  F=1,m_{f}=-1\right\rangle $ (denoting $\left|  1\right\rangle $ for
short) and $\left|  F=2,m_{f}=1\right\rangle $ ($\left|  2\right\rangle $)
coupled by a radiant field in a single trap which applies a potential $V_{l}$
($l=1,2$) to the atoms of $l$-state. The interaction between atoms is
considered as elastic two-body collision with the $\delta-$function-type
potential. In the formalism of second quantization, Hamiltonian of this system
with a quantized field reads%

\begin{equation}
H=\sum_{l=1,2}H_{l}+H_{int}+H_{ed}+H_{R} \label{1}%
\end{equation}
with%

\begin{align*}
H_{l}  &  =\int d^{3}\mathbf{r}\{\Psi_{l}^{+}(\mathbf{r})[-\frac{\hbar^{2}%
}{2m}\Delta^{2}+V_{l}(\mathbf{r})]\Psi_{l}(\mathbf{r})+\\
&  \frac{q_{l}}{2}\Psi_{l}^{+}(\mathbf{r})\Psi_{l}^{+}(\mathbf{r})\Psi
_{l}(\mathbf{r})\Psi_{l}(\mathbf{r})\},
\end{align*}

\[
H_{int}=q_{1,2}\int d^{3}\mathbf{r}\Psi_{1}^{+}(\mathbf{r})\Psi_{2}%
^{+}(\mathbf{r})\Psi_{1}(\mathbf{r})\Psi_{2}(\mathbf{r}),
\]

\[
H_{f}=i\hbar\Omega\int d^{3}\mathbf{r}[\Psi_{1}^{+}(\mathbf{r})\Psi
_{2}(\mathbf{r})+\Psi_{2}^{+}(\mathbf{r})\Psi_{1}(\mathbf{r}%
)](ae^{i\mathbf{k\cdot r}}-e^{-i\mathbf{k\cdot r}}a^{+}),
\]

\[
H_{R}=\omega a^{+}a,
\]
where $\Psi_{l}(\mathbf{r})$ is boson field operator of atoms in the internal
atomic state-$l$. $q_{l}=\ 4\pi\hbar^{2}\rho_{l}/m$ measures the interaction
between atoms of the same species while $q_{1,2}=4\pi\hbar^{2}\rho_{1,2}/m$
denotes the atom-interaction of different species, where $\rho_{l}$ and
$\rho_{1,2}=\rho_{2,1}$ are the intraspecies $s-$wave scattering lengths and
can be tuned by an external magnetic field via Feshbach resonance \cite{26}.
$a$, $a^{+}$ denote respectively the radiation-field annihilation and creation
operators with frequency $\omega$. $H_{f}$ describes the electric dipole
transition between two-species atoms induced by the radiation-field in the
Schr$\ddot{o}$dinger picture \cite{27}. The interaction constant $\Omega$ is
known as the Rabi frequency and $\mathbf{k}$ denotes the wave vector of the
field. It should be noticed that the usual rotating wave approximation is not
needed in our formulation of the interaction between atom and field.

In the single-mode approximation of condensate such that $\Psi_{1}%
(\mathbf{r})=c_{1}\phi_{1}(\mathbf{r}),\Psi_{2}(\mathbf{r})=c_{2}\phi
_{2}(\mathbf{r}),$ where $c_{1}$ and $c_{2}$ are the annihilation operators,
the Hamiltonian (1) becomes
\begin{align}
H  &  =\sum_{l=1,2}(\omega_{l}c_{l}^{+}c_{l}+\frac{\eta_{l}}2c_{l}^{+}%
c_{l}^{+}c_{l}c_{l})+\chi c_{1}^{+}c_{1}c_{2}^{+}c_{2}+\label{2}\\
&  \lambda_{\Omega}(c_{1}^{+}c_{2}+c_{2}^{+}c_{1})(a+a^{+})+\omega
a^{+}a,\nonumber
\end{align}
where $\omega_{l}=\int d^{3}\mathbf{r}\{\phi_{l}^{*}(\mathbf{r})[-\frac
{\hbar^{2}}{2m}\Delta^{2}+V_{l}(\mathbf{r})]\phi_{l}(\mathbf{r})$, $\eta
_{l}=q_{l}\int d^{3}\mathbf{r}\left|  \phi_{l}(\mathbf{r})\right|  ^{4}$,
$\chi=q_{1,2}\int d^{3}\mathbf{r}\left|  \phi_{1}(\mathbf{r})\right|
^{2}\left|  \phi_{2}(\mathbf{r})\right|  ^{2}$ and $\lambda_{\Omega}%
=-\hbar\Omega\int d^{3}\mathbf{r}\phi_{1}^{*}(\mathbf{r})\sin(\mathbf{k\cdot
r)}\phi_{2}(\mathbf{r})$. Hamiltonian (2), which is the starting point of the
paper, is a full quantum description of atom-condensate interacting with the
radiant field. In terms of the pseudoangular momentum operators with the
Schwinger relations defined as $S_{x}=\frac12(c_{1}^{+}c_{2}+c_{2}^{+}%
c_{1}),\quad S_{y}=\frac1{2i}(c_{1}^{+}c_{2}-c_{2}^{+}c_{1}),$and$\quad
S_{z}=\frac12(c_{1}^{+}c_{1}-c_{2}^{+}c_{2}),$ where the Casimir invariant is
$S^{2}=\frac N2(\frac N2+1)$ with $N$ being the total atom number, Hamiltonian
(2) is converted to $H=\omega a^{+}a+\omega_{0}S_{z}+qS_{z}^{2}+2\lambda
_{\Omega}S_{x}(a^{+}+a)$ with $\omega_{0}=\omega_{1}-\omega_{2}+(N-1)(\eta
_{2}-\eta_{1})/2$, $q=[(\eta_{1}+\eta_{2})/2-\chi]$, which can be rewritten
apart from a trivial constant as
\begin{equation}
H=\omega a^{+}a+\tilde{\omega}_{0}S_{z}+\frac\lambda{\sqrt{N}}(S_{+}%
+S_{-})(a^{+}+a)+\frac\nu NS_{+}S_{-}, \label{3}%
\end{equation}
where $S_{\pm}=S_{x}\pm iS_{y}$ , $\nu=-Nq$, $\lambda=\sqrt{N}$ $\lambda
_{\Omega}$ and $\tilde{\omega}_{0}=\omega_{0}+q\simeq\omega_{0}$ since the
parameter $q$ is a negligibly small number in our case seen below. The
prefactor $1/N$ makes a finite free energy per atom in the thermodynamic limit
$N\rightarrow\infty$.

We first of all derive the ground state energy spectrum as a function of the
coupling parameter $\lambda$ in order to reveal the QPT. To this end we use
the Holstein-Primakoff transformation (HPT) of the angular momentum operators
\cite{28} $S_{+}=b^{+}\sqrt{N-b^{+}b}$, $S_{-}=\sqrt{N-b^{+}b}b$ and
$S_{z}=(b^{+}b-N/2)$, where $[b,b^{+}]=1$, to rewrite the Hamiltonian and
subsequently introduce shifting boson operators $\ \widetilde{a}^{+}\ $and
$\widetilde{b}^{+}\ $with properly scaled auxiliary parameters $\alpha$ and
$\beta$ such that $\widetilde{a}^{+}=a^{+}+\sqrt{N}\alpha$ and $\widetilde
{b}^{+}=b^{+}-\sqrt{N}\beta$ \cite{29}. We then expand the Hamiltonian
expressed with the new boson operators $\widetilde{a}^{+}$ and $\widetilde
{b}^{+}$ as power series of $1/\sqrt{N}$ and obtain $H=NH_{0}+N^{1/2}%
H_{1}+N^{0}H_{2}+\cdot\cdot\cdot$, the first term of which gives rise to the
Hartree-Bogoliubov ground state energy \cite{30}%

\begin{equation}
E_{0}=\left\{
\begin{array}
[c]{l}%
-N\omega_{0}/2\text{ ,\ \ \ \ \ \ \ \ \ \ \ \ \ \ \ \ \ \ \ \ \ \ \ \ }%
\lambda<\lambda_{c}\\
-N[(\frac{\lambda^{2}}\omega-\frac\nu4)(1-\delta^{2})+\frac{\omega_{0}\delta
}2]\ ,\text{ }\lambda>\lambda_{c}%
\end{array}
\right.  , \label{4}%
\end{equation}
where $\delta=\omega\omega_{0}/(4\lambda^{2}-\omega\nu)$ and%

\begin{equation}
\lambda_{c}=\frac12\sqrt{\omega(\omega_{0}+\nu)}\text{ or }\lambda_{\Omega
,c}=\frac12\sqrt{\omega(\omega_{0}+\nu)/N}. \label{5}%
\end{equation}
The auxiliary parameters $\alpha$ and $\beta$ are determined from minimizing
the ground state energy and thus are given by%

\begin{align}
\alpha &  =\left\{
\begin{array}
[c]{l}%
0\text{ \ \ \ \ \ \ \ \ \ \ \ \ , }\lambda<\lambda_{c}\\
\frac\lambda\omega\sqrt{1-\delta^{2}}\text{ \ , }\lambda>\lambda_{c}%
\end{array}
\right.  \text{, }\nonumber\\
\beta &  =\left\{
\begin{array}
[c]{l}%
0\text{ \ \ \ \ \ \ \ \ \ , }\lambda<\lambda_{c}\\
\sqrt{(1-\delta)/2}\text{\ \ \ \ \ \ \ \ \ , }\lambda<\lambda_{c}%
\end{array}
\right.  .
\end{align}
The ground state energy as a function of the coupling parameter is shown in
Fig. 1 which indicates the typical second-order phase transition at the
transition point $\lambda_{c}$. The QPT is similar to that observed in the
Dicke Hamiltonian \cite{25} and the two phases may be respectively called the
normal ($\lambda<\lambda_{c}$) and super-radiant ($\lambda>\lambda_{c}$)
phases according to the properties of the ground state and the atom population
imbalance shown below.

Having determined the transition point $\lambda_{c}$ we now study the critical
behavior of the system at the phase transition point. For $\lambda<\lambda
_{c}$ , $\alpha=\beta=0$, we may take the approximation\textbf{\ }%
$\sqrt{N-b^{+}b}\simeq\sqrt{N}$\textbf{\ }in the HPT of the angular momentum
operators and the effective Hamiltonian is seen to be $H_{<}=\omega
a^{+}a+(\omega_{0}+\nu)b^{+}b+\lambda(b+b^{+})(a^{+}+a)-\frac{N\omega_{0}}%
2,$which is bilinear in the bosonic operators and can thus be diagonalized in
general with the Bogoliubov transformation. Here we adopt an alternative
approach in terms of the\ canonical operators $a^{+}=(\sqrt{\omega}%
x-ip_{x}/\sqrt{\omega})/\sqrt{2}$ , $b^{+}=(\sqrt{\omega_{0}+\nu}%
y-ip_{y}/\sqrt{\omega_{0}+\nu})/\sqrt{2}$\ and the Hamiltonian, in a rotating
coordinate system such that $x=q_{1}\cos\gamma+q_{2}\sin\gamma,$ $y=-q_{1}%
\sin\gamma+q_{2}\cos\gamma$ with $\gamma=\frac12\arctan4\lambda\sqrt
{\omega(\omega_{0}+\nu)}/[(\omega_{0}+\nu)^{2}-\omega^{2}],$ has the simple form%

\begin{equation}
H_{(<)}=\frac{1}{2}[E_{(<)-}^{2}q_{1}^{2}+p_{1}^{2}+E_{(<)+}^{2}q_{2}%
^{2}+p_{2}^{2}-\omega_{0}-\omega-\nu]-\frac{N\omega_{0}}{2}, \label{7}%
\end{equation}
where the collective modes of fundamental excitations
\begin{equation}
E_{(<)\pm}^{2}=\frac{1}{2}\{[\omega^{2}+(\omega_{0}+\nu)^{2}]\pm\sqrt
{\lbrack\omega^{2}-(\omega_{0}+\nu)^{2}]^{2}+16\lambda^{2}\omega(\omega
_{0}+\nu)}\}
\end{equation}
are recognized as the atomic and photonic modes respectively. The $\lambda
-$dependence property of photonic mode $E_{(<)-}$ is shown in Fig.2. When the
effective coupling strength $\lambda$ approaches the critical value
$\lambda_{c}$, the photonic mode $E_{(<)-}$ vanishes as
\begin{equation}
E_{(<)-}(\lambda\rightarrow\lambda_{c})\sim\sqrt{\frac{32\lambda_{c}^{3}%
\omega^{2}}{16\lambda_{c}^{4}+\omega^{4}}}\left\vert \lambda-\lambda
_{c}\right\vert ^{\frac{1}{2}}. \label{9}%
\end{equation}

Above the phase transition point ( $\lambda>\lambda_{c}$) a similar effective
Hamiltonian in terms of canonical operators $P$ and $Q$ is found as%

\begin{align}
H_{(>)}=  &  \frac{1}{2}[E_{(>)-}^{2}Q_{1}^{2}+P_{1}^{2}+E_{(>)+}^{2}Q_{2}%
^{2}+P_{2}^{2}-\omega_{0}-\omega_{1}]\label{10}\\
&  -N[(\frac{\lambda^{2}}{\omega}-\frac{\nu}{4})(1-\delta^{2})+\frac
{\omega_{0}\delta}{2}]\nonumber
\end{align}
with the corresponding atomic and photonic modes
\begin{align}
E_{(>)\pm}^{2}  &  =\frac{1}{2}\{[\omega^{2}+\frac{\omega_{1}^{2}\ (\omega
_{1}+2\omega_{3}+4\omega_{4})}{\omega_{1}-2\omega_{3}}]\nonumber\\
&  \pm\sqrt{\lbrack\omega^{2}-\frac{\omega_{1}^{2}\ (\omega_{1}+2\omega
_{3}+4\omega_{4})}{\omega_{1}-2\omega_{3}}]^{2}+\frac{16\omega_{1}^{2}%
\omega_{2}^{2}\omega}{\omega_{1}-2\omega_{3}}}\}
\end{align}
(see Fig.2) where $\omega_{1}=\frac{4\lambda^{2}}{\omega}-\nu(1-\delta)$,
$\omega_{2}=\lambda\delta\sqrt{\frac{2}{(1+\delta)}}$, $\omega_{3}%
=\frac{\lambda^{2}}{\omega}(1-\delta)-\frac{\nu}{2}(1-\delta)$, $\omega
_{4}=\frac{\lambda^{2}}{\omega}\frac{(1-\delta)^{2}}{2(1+\delta)}$. The
scaling behavior of the photonic mode $E_{(>)-}$ at the critical point
$\lambda_{c}$ is the same as that in the normal phase as it should be. The
scaling behavior of the photonic mode, $E_{-}\varpropto\left\vert
\lambda-\lambda_{c}\right\vert ^{\frac{1}{2}}$ , with the exponent describing
a characteristic divergence length $\xi=(E_{-})^{-1/2}$ \cite{19}. The energy
of the atomic mode $E_{+}$ tends to a value of $\sqrt{\omega^{2}+(\omega
_{0}+\nu)^{2}}$ as $\lambda\rightarrow\lambda_{c}$ from both sides.

The atom population imbalance between two internal states of $^{87}$Rb atoms
in the ground state, $\Delta N=N_{1}-N_{2}$ $=2\left\langle S_{z}\right\rangle
$, is found as%

\begin{equation}
\frac{\Delta N}N=\left\{
\begin{array}
[c]{l}%
-1,\text{ \ }\ \ \ \ \ \ \ \ \ \ \ \text{ }\lambda<\lambda_{c}\\
-\omega\omega_{0}/(4\lambda^{2}-\omega\nu),\ \lambda>\lambda_{c}%
\end{array}
\right.  ,
\end{equation}
which\ as a function of the coupling strength $\lambda$ is plotted in Fig.(3).
In the normal phase ( $\lambda<\lambda_{c}$) atomic ensemble is essentially in
the lower energy state, whereas acquires macroscopic excitation above the
transition point. This QPT may be observed by measuring the abrupt change of
the atom population imbalance over increasing the Rabi frequency in practical
experiments. In practice the abrupt change of the atom population imbalance
can be measured experimentally by the variation of absorption spectrum of the
light with respect to the coupling parameter $\lambda_{\Omega_{c}}$. Then a
critical point is to see whether or not the critical value of coupling
parameter $\lambda_{\Omega_{c}}=\frac{\lambda_{c}}{\sqrt{N}}$ is realizable in
BEC experiments of $^{87}$Rb atoms. For the magnetic trap of an axial
frequency $\omega_{z}=69$ Hz and radial frequency $\omega_{x,y}=21$ Hz, the
$s-$wave scattering lengths are $\rho_{1}=5.36$ nm, $\rho_{2}=5.70$ nm,
$\rho_{1,2}=\rho_{2,1}=5.53$ nm, and the number of trapped atoms is
$N=5\times10^{5}$ \cite{12}. The parameter $\omega_{0}$ is found approximately
as $\omega_{0}=4.1\times10^{2}$ Hz. The laser frequency used by Ketterle'group
for Rayleigh scattering is $\omega=1.7$ GHz \cite{15}. With these experimental
parameters the energy scale of the critical coupling parameter at transition
point is evaluated as $\lambda_{\Omega,c}= $ $4.47$ nk, which is far below the
critical temperature $T_{c}=150$ nk of dilute $^{87}$Rb gas BEC and thus the
quantum phase transition is experimentally realizable.

In conclusion, we have demonstrated a second-order quantum phase transition in
two-species BECs coupled by a radiant field with Raman coupling similar to the
well known transition in the Dicke model of quantum optics where there is no
atom-atom interaction. The quantum phase transition is more natural in BEC and
can be observed directly by measuring the atom population imbalance between
the two components of the BECs.

One of authors (G.C.) thanks Dr. EMARY for helpful discussions and valuable
suggestions. This work was supported by the Natural Science Foundation of
China under Grant Nos.10475053, 90403034, 90406017, 60525417,and by the
Natural Science Foundation of Zhejiang Province under Grant No.Y605037. W.M.L.
also acknowledges the support from CAS with the project: International Team on
Superconductivity and Novel Electronic Materials (ITSNEM).

\end{document}